\documentclass[sort&compress,final,3p,times,twocolumn,fixfloat]{elsarticle}

\usepackage{graphicx}% eps fig, package for more complicated commands
\usepackage{amssymb} % various useful mathematical symbols
\usepackage{amsmath} % extended theorem environments
\usepackage{times}
\usepackage{tabularx}

\usepackage{txfonts}
\usepackage{lineno}
\journal{Physics Letters B}

 \usepackage{ifpdf}
 \ifpdf
 \usepackage[pdftex]{hyperref}
 \else
 \usepackage[hypertex]{hyperref}
 \fi

 \hypersetup{
   pdftitle={},%
   pdfauthor={},%
   pdfsubject={},%
   pdfkeywords={},%
   pdfstartview={},%
   bookmarksopen=true, breaklinks=true, debug=true, %
   colorlinks=true, linkcolor=red, citecolor=blue, urlcolor=blue
 }

\usepackage{float} % improved interface for floating objects
\usepackage[caption=false]{subfig} % supersedes the subfigure package
\newcount\savefnused
\newcount\savefndone

\newcommand{\savefootnote}[2][\empty]% #1=number (optional), #2=text
{\ifx\empty#1\footnotemark\else\footnotemark[#1]\fi
 \global\advance\savefnused by 1
 \expandafter\xdef\csname savefnmark\the\savefnused\endcsname{\thefootnote}%
 \expandafter\xdef\csname savefntext\the\savefnused\endcsname{#2}%
}
\newcommand{\flushfootnote}{\loop\ifnum\savefndone<\savefnused
  \global\advance\savefndone by 1
  \footnotetext[\csname savefnmark\the\savefndone\endcsname]%
    {\csname savefntext\the\savefndone\endcsname}%
  \global\expandafter\let\csname savefnmark\the\savefndone\endcsname\relax
  \global\expandafter\let\csname savefntext\the\savefndone\endcsname\relax
\repeat}

\usepackage{multirow,bigdelim}

\usepackage{textcomp}
\usepackage{comment}

\usepackage{lineno}
\usepackage{float} % improved interface for floating objects
\usepackage{subfig} % supersedes the subfloat package
\usepackage{amsmath} %
\usepackage{amstext}
\usepackage{graphicx}%
\usepackage{tabularx,ragged2e}%

\newcolumntype{Y}{>{\centering\arraybackslash}X}

\def\GeV{{\rm GeV}}
\def\GeV2{{\rm GeV}^2}

\begin{document} 
\begin{frontmatter}

\title{Dark matter scattering cross section and dynamics in dark Yang-Mills theory}
\author[a,b]{Nodoka Yamanaka}
\author[c]{Hideaki Iida}
\author[d,e,f]{Atsushi Nakamura}
\author[g,h,i,e]{Masayuki Wakayama}
\address[a]{Yukawa Institute for Theoretical Physics, Kyoto University, Kitashirakawa-Oiwake, Kyoto 606-8502, Japan}
\address[b]{Amherst Center for Fundamental Interactions, Department of Physics, University of Massachusetts Amherst, MA 01003, USA}
\address[c]{Department of Physics, The University of Tokyo, 7-3-1 Hongo, Bunkyo-ku, Tokyo 113-0033, Japan}
\address[d]{Pacific Quantum Center, Far Eastern Federal University, Sukhanova 8, Vladivostok, 690950, Russia}
\address[e]{Research Center for Nuclear Physics, Osaka University, Ibaraki, Osaka 567-0047, Japan}
\address[f]{Theoretical Research Division, Nishina Center, RIKEN, Wako, Saitama 351-0198, Japan}
\address[g]{School of Science and Engineering, Kokushikan University, Tokyo 154-8515, Japan}
\address[h]{Center for Extreme Nuclear Matters (CENuM), Korea University, Seoul 02841, Republic of Korea}
\address[i]{Department of Physics, Pukyong National University (PKNU), Busan 48513, Republic of Korea}

\begin{abstract}
{\small
We calculate for the first time the scattering cross section between lightest glueballs in $SU(2)$ pure Yang-Mills theory, which are good candidates of dark matter.
In the first step, we evaluate the interglueball potential on lattice using the HAL QCD method, with several lattice spacings ($\beta = 2.1, 2.2, 2.3, 2.4$, and 2.5).
The systematics associated with nonzero angular momentum effect is removed by subtracting the centrifugal force.
The statistical accuracy is improved by employing the cluster-decomposition error reduction technique and by using all space-time symmetries.
We then determine the low energy glueball effective Lagrangian and the scattering cross section at low energy, which is compared with the observational constraint on the dark matter self-scattering.
We derive the lower bound on the scale parameter of the $SU(2)$ Yang-Mills theory, as $\Lambda > 60$ MeV.
}
\end{abstract}

\date{\today}
\end{frontmatter}
%\pacs{98.80.-k,95.35.+d,11.15.-q,12.39.Mk}
%Cosmology, 98.80.-k
%Dark matter, 95.35.+d
%Gauge field theories, 11.15.-q
%Glueball and nonstandard multi-quark/gluon states, 12.39.Mk
%

%%%%%%%%%%%%%%%%%%%%%%%%%%%%%%%%%%%%%%%%%%%%%%%%%%%%%%%%%%%%%%%%%%%%%%%%%%%%

The existence of a significant amount of dark matter (DM) \cite{Bertone:2004pz,Munshi:2006fn,Arcadi:2017kky,Battaglieri:2017aum} in our Universe is supported by many physical data.
Its presence was first suggested by the observations of the galactic rotation curve \cite{Zwicky:1933gu,Davis:1985rj}, and is now made firm by that of the density profile of the bullet clusters \cite{Clowe:2006eq}.
The quantity of DM is nowadays very precisely known thanks to the progress of the observation of the cosmic microwave background, with 27\% of the energy composition of our Universe \cite{Ade:2015xua}.
Another important feature is the necessity of the DM in the formation of the large scale structure of the Universe \cite{Blumenthal:1984bp,Navarro:1995iw,Navarro:1996gj,Oguri:2011dt,Umetsu:2015baa}.
Owing to the above strong arguments, the study of the DM is one of the most essential subjects of fundamental physics.

Thanks to the progress of the observations of the gravitational microlensing, the DM of astrophysical compact objects are excluded in a very wide mass region \cite{Niikura:2017zjd,Sasaki:2018dmp}.
This suggests that the DM is mainly composed of particles which interact very weakly with the visible sector \cite{Roszkowski:2017nbc}. 
It is known that the standard model (SM) does not contain any fields which fulfill the properties of DM, and many model candidates are under investigation \cite{Bertone:2004pz,Arcadi:2017kky,Battaglieri:2017aum,Roszkowski:2017nbc}.
Among these new physics, the most popular ones are extensions of the SM with new sector(s), often protected by discrete symmetries. 
However, these theories often have problems with the results of direct \cite{Aprile:2012nq,Agnese:2013rvf,Cui:2017nnn,Crisler:2018gci} and indirect \cite{ArkaniHamed:2008qn,TheFermi-LAT:2015kwa,Aartsen:2016ngq,Aguilar:2016kjl,Aguilar:2019owu} detection experiments as well as with search using colliders \cite{Kahlhoefer:2017dnp,Aaboud:2017phn,Sirunyan:2018xlo}, due to the relatively large coupling with the SM required to keep consistency with the relic density (the so-called ``WIMP miracle'') \cite{Jungman:1995df,Chu:2011be}.

As opposed to the above ``elementary'' DM particles, we have the composite DM \cite{Cline:2013zca,Foot:2014uba,Kribs:2016cew,Hertzberg:2019bvt}, which has several classes.
Here we investigate the less discussed but the most natural DM model, the dark $SU(N_c)$ Yang-Mills theory (YMT)
\begin{equation}
{\cal L}_{\rm YM}
=
-\frac{1}{4} \sum_{a=1}^{N_c^2-1} F^{\mu \nu}_a F_{\mu \nu}^a
,
\label{eq:YMT}
\end{equation}
in which the lightest glueballs are candidates of DM \cite{Carlson:1992fn,Faraggi:2000pv,Soni:2016gzf,Forestell:2016qhc,Soni:2016yes,Forestell:2017wov,Kang:2019izi}.
We emphasize that the above gauge theory is a stand alone YMT, and the glueballs will not decay to SM particles.
The attractive features of this scenario are as follows.
First, the scale parameter $\Lambda$ is only controlled by the color number $N_c$ through the dimensional transmutation, and the theory is thus almost free from the hierarchical problem.
Second, it is also part of another very natural scenario where vectorlike quarks are present in the Lagrangian, but with masses heavier than $\Lambda$.
Finally, it may naturally be embedded in more ultraviolet complete frameworks such as the grand unification or string theory \cite{Carlson:1992fn,Dienes:1996du,Benakli:1998ut,Faraggi:2000pv}.

The glueball is a gluonic bound state \cite{Jaffe:1975fd,Cornwall:1981zr,Klempt:2007cp,Mathieu:2008me} which purely reflects the nonperturbative physics of nonabelian gauge theory, and it cannot be investigated perturbatively.
Indeed, it has been an excellent target for nonperturbative methodologies such as the lattice gauge theory \cite{Azcoiti:1983ab,deForcrand:1984eeq,Albanese:1987ds,Bali:1993fb,Sexton:1995kd,Teper:1998kw,Morningstar:1999rf,Ishii:2001zq,Ishii:2002ww,Lucini:2004my,Chen:2005mg,Lucini:2010nv,Gregory:2012hu,Lucini:2012gg} or holographic approaches \cite{Csaki:1998qr}, and its mass spectrum has extensively been evaluated.
Despite these tremendous works, the properties of the glueballs are still obscure.
In QCD, the observation of glueballs is not yet conclusive, although there are several candidates such as $f_0$(1500) and $f_0$(1710), and they are currently actively searched in experiments \cite{Crede:2008vw}.
The determination of the glueball is essentially difficult due to the mixing with other hadronic states \cite{Cheng:2006hu,Ochs:2013gi}, but this issue is absent in the case of the dark YMT, since the glueball is the lightest state and there is no mixing with other particles.

In considering the particle DM scenario and its cosmology, inevitable topics to be discussed are multi-particle processes, such as the self-interaction (or scattering) among DM or the production of relic DM.
The self-interaction is especially important since it is bound by observations such as the density profile of the galactic halo or the collision between galaxies.
As for the scale smaller than kpc, several phenomena, which were thought to be ``problems'' in relation to the density profile, are known to be explained with finite DM self-interaction \cite{Carlson:1992fn,Spergel:1999mh,Vogelsberger:2012ku,Rocha:2012jg,Zavala:2012us,Peter:2012jh,Tulin:2013teo,Tulin:2017ara}, although these topics are still a matter of controversy \cite{Chan:2015tna,Wetzel:2016wro,Kim:2017iwr,Bullock:2017xww,vanDokkum:2018vup}.
A promising approach for the determination of the dark YMT is then to derive the relation between the glueball scattering cross section and the scale parameter $\Lambda$, then constrain the latter from observational data.
Currently, this relation is not known from the first principles although it was challenged in model calculations \cite{daSilva:2004rm,daSilva:2006avt,daSilva:2012sb,Novello:2012kta,Soni:2016gzf,Soni:2016yes}.

The quantification of the glueball scattering will further serve us to derive the low energy $0^{++}$ glueball effective field theory which respects the conformal Ward identity \cite{Schechter:1980ak,Migdal:1982jp}
\begin{equation}
{\cal L}_{\phi}
=
\frac{1}{2} e^{2\frac{\phi}{f_\phi}} (\partial^\mu \phi ) (\partial_\mu \phi )
-H_0 
\left(
\frac{1}{4} - \frac{\phi}{f_\phi}
\right)
e^{4\frac{\phi}{f_\phi}}
,
\label{eq:glueballEFT}
\end{equation}
where the decay constant $f_\phi$ and the vacuum expectation value $H_0$ are the only two free parameters.
This means that it is actually possible to completely determine the low energy dynamics of the YMT by using the mass and the glueball cross section.
After the determination of the above Lagrangian, it may then be used to calculate and thus predict other observables such as the relic density.
The analysis of the glueball scattering on lattice is therefore the key in the quantification of the cosmology of dark YMT.

The lattice gauge theory simulation, which is currently almost the only way to systematically calculate observables related to glueballs, could so far only quantify their masses \cite{Albanese:1987ds,Bali:1993fb,Teper:1998kw,Morningstar:1999rf,Ishii:2001zq,Ishii:2002ww,Lucini:2004my,Chen:2005mg,Lucini:2010nv,Gregory:2012hu}.
Recently, however, there have been drastic improvements in the calculations of the multi-hadron systems on lattice.
The way to calculate the scattering phase shift of two particles in a finite volume lattice was first conceived by L\"{u}scher \cite{Luscher:1990ux}.
This method was then successfully implemented in numerical calculations, and results for several systems are currently available \cite{Beane:2010em}.
As an alternative approach, we also have the HAL QCD method in which the phase shift is indirectly extracted via the potential \cite{Ishii:2006ec,Aoki:2009ji,HALQCD:2012aa,Miyamoto:2019jjc}.
This approach is known to be able to extract the interhadron potential without waiting for the formation of the plateau, which greatly reduces the computational cost.

In this letter, we discuss the self-interaction of the DM in the YMT by calculating the interglueball scattering cross section in $SU(2)$ lattice gauge theory employing the HAL QCD method.
Our goal is, through the calculation of the cross section, to constrain the scale parameter of the $SU(2)$ YMT from observations and to determine the low energy $0^{++}$ glueball effective Lagrangian.
We also extrapolate our results to $N_c \ge 3$ using the large $N_c$ argument.
The full discussion can be found in another paper \cite{Yamanaka:2019yek}.

To proceed, we simulate $N_c=2$ pure YMT (\ref{eq:YMT}) on lattice, with $\beta = 2.1, 2.2, 2.3, 2.4$, and 2.5.
The scale of the lattice is expressed in units of the scale parameter $\Lambda$ which is unknown, since the mass and other dimensionful quantities related to the DM are not known.
The result of our work will actually quantitatively relate them with $\Lambda$.
The relation between the lattice spacing $a$ and $\Lambda$ was fitted through the calculation of the string tension $\sigma$.
For the general $N_c$, we have \cite{Allton:2008ty,Teper:2009uf}
\begin{eqnarray}
\frac{\Lambda}{\sqrt{\sigma}}
=
0.503(2)(40)+ \frac{0.33(3)(3)}{N_c^2}
.
\end{eqnarray}
The values of $a\sqrt{\sigma}$ were taken from Refs. \cite{Teper:1997tq,Teper:1998kw} (e.g. $a = 0.107(8) \Lambda^{-1}$ for $\beta = 2.5$). 
We used the pseudo-heat bath method to generate gauge configurations.

We now define the $0^{++}$ glueball operator:
\begin{eqnarray}
\hspace{-2.0em}
\phi (t, \vec{x}) 
&=&
{\rm Re} [
P_{12} (t, \vec{x}) 
+P_{12} (t, \vec{x}+a\vec{e}_3) 
+P_{23} (t, \vec{x}) 
+
\nonumber\\
&&
P_{23} (t, \vec{x}+a\vec{e}_1) 
+P_{31} (t, \vec{x}) 
+P_{31} (t, \vec{x}+a\vec{e}_2) 
]
,
\label{eq:glueballop}
\end{eqnarray}
where $P_{ij}$ are the plaquette operator in $i-j$ direction, and $\vec{e}_{1,2,3}$ are the unit vectors.
Since the $0^{++}$ glueball has the same quantum number as the vacuum, we have to subtract its expectation value as $\tilde \phi (t, \vec{x}) \equiv \phi (t, \vec{x}) - \langle \phi (t, \vec{x}) \rangle$ in order to calculate the physical glueball correlators.
We may also improve the glueball operator using the APE smearing \cite{Albanese:1987ds,Ishii:2001zq,Ishii:2002ww}.
With this optimization, we obtained the glueball mass $m_\phi = 6.32(46) \Lambda$ at $\beta = 2.5$.

\begin{figure}[htb]
\includegraphics[width=8.5cm]{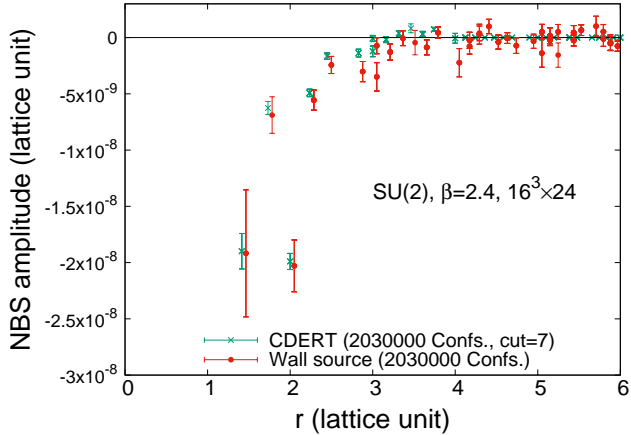}
\caption{\label{fig:BS}
Glueball NBS amplitude (\ref{eq:NBS}) with 1-body source (${\cal J} = \tilde \phi$) measured at $t=2$ in a $16^3 \times 24$ lattice with $\beta = 2.4$.
We compare the improvement of the signal thanks to the CDERT with the cutoff $r\le 7$.
The characteristic oscillation is due to the effect of nonzero angular momenta (centrifugal force) of lattice discretization.
}
\end{figure}

The physical information of the scattering between two hadrons can be extracted from the Nambu-Bethe-Salpeter (NBS) amplitude.
For the glueball two-body scattering, it is defined as follows:
\begin{equation}
\Psi_{\phi \phi} 
(t,\vec{x}-\vec{y})
\equiv
\frac{1}{V} \sum_{\vec{r}} 
\langle 0 | T[\tilde \phi (t, \vec{x}+\vec{r})\tilde \phi (t, \vec{y}+\vec{r}) {\cal J}(0)] | 0 \rangle
,
\label{eq:NBS}
\end{equation}
where ${\cal J}$ is the source operator with the same quantum number as the two-glueball state.
Here two important features have to be mentioned.
First, in the case of the glueball, ${\cal J}$ may be chosen as an $n$-body operator ($\tilde \phi^n$) with arbitrary positive-definite integer $n$, due to the $0^{++}$ quantum number.
Second, the multi-glueball operators also have expectation values which have to be subtracted.
The correlator (\ref{eq:NBS}) is purely gluonic, and the statistical error is significant in the lattice calculation.
To improve the signal, we use all space-time symmetries (space-time translations and cubic rotations) to effectively increase the statistics.
For $\beta=2.4$, we also employ the cluster-decomposition error reduction technique (CDERT) \cite{Liu:2017man} to remove the fluctuation of the vacuum insertion.
In our work, we used the cut $r\le 7$ (lattice unit) for which the systematic error is found to be less than the statistical one (see Fig.~\ref{fig:BS}). 
Thanks to the CDERT, we could reduce the error bar by more than twice, which demonstrates its efficacy.

Let us now extract the scattering phase shift.
The direct way to calculate it is to Fourier transform the NBS amplitude and inspect the momentum modulation of the energy (so-called L\"{u}cher's method) \cite{Luscher:1990ux}.
This approach was successful in the mesonic sector \cite{Beane:2010em}, but in the case of the glueball, we encounter a problem, due to the necessity of taking the plateau of the NBS amplitude.
Indeed, the NBS amplitude mixes with the single glueball (2-point) correlator so that taking the plateau will always show the single glueball mass as the energy of the system.
We might, of course, remove the one-glueball state by diagonalizing the NBS amplitude, but the glueball spectrum has other resonances close to the two-body threshold, so the extraction of the two-glueball scattering in this approach is highly challenging.
An alternative approach to calculate the scattering phase shift is to indirectly extract it via the potential (HAL QCD method) \cite{Aoki:2009ji}.
This method has the crucial advantage that we do not need the ground state saturation for obtaining the potential \cite{HALQCD:2012aa}.
In particular, the glueball correlators are in general very noisy, so the use of this method is almost mandatory if one wants to keep good statistical accuracy.
In addition, the potential handled in the HAL QCD method does not depend on the renormalization scale \cite{Ishii:2006ec,Aoki:2009ji}, which will allow us to connect the results of the calculations with different $\beta$ without special corrections.
We however have to keep in mind that the potential is not an observable, and it may depend on the choice of the operators.

Let us now describe the formalism of the calculation of the interglueball potential on lattice.
The nonlocal potential $U({\vec r},{\vec r}')$ is extracted according to the following time-dependent Schr\"{o}dinger-like equation \cite{HALQCD:2012aa}
\begin{eqnarray}
&&
\biggl[
\frac{1}{4m_\phi} \frac{\partial^2}{\partial t^2}-\frac{\partial}{\partial t} + \frac{1}{m_\phi} \nabla^2
-\frac{(\vec{r} \times \vec{\nabla})^2}{m_\phi r^2}
\biggr]
R(t,{\vec r})
\nonumber\\
&&
=
\int d^3r' U({\vec r},{\vec r}')R(t,{\vec r}')
,
\label{eq:HALQCD}
\end{eqnarray}
where $R(t,{\vec r}) \equiv \frac{\Psi_{\phi \phi} (t,{\vec r}) }{e^{-2m_\phi t}}$.
The last term corresponds to the subtraction of the centrifugal force, and we expect to remove systematic effects of nonzero angular momentum contribution (see the oscillation of the NBS amplitude of Fig. \ref{fig:BS}).
Here $t$ should be chosen so that $1/t$ is less than the inelastic threshold $ m_\phi (=3m_\phi - 2m_\phi )$.
In our calculations with $\beta \ge 2.4$ $(\le 2.3)$, it is enough to take the data from $t =2$ $(=1)$.
Regarding $\beta =2.1$ and 2.2, we use the time-independent HAL QCD method [extraction without using the time derivative terms in Eq. (\ref{eq:HALQCD})] since the discretization error in the time direction is too important to be used in the time-dependent formalism.
The physics addressed here is nonrelativistic, and the potential should therefore be to a good approximation local and central $U({\vec r},{\vec r}') \approx V_{\phi \phi} (\vec{r}) \delta (\vec{r}-\vec{r}')$.

\begin{figure}[htb]
\includegraphics[width=8.5cm]{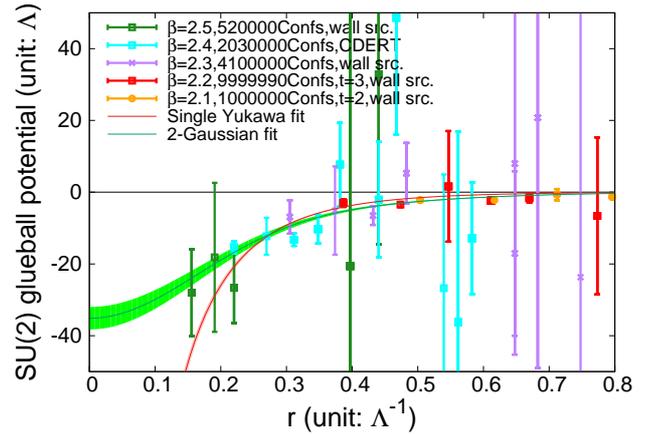}
\caption{\label{fig:potential}
Interglueball potential calculated on lattice in the $SU(2)$ YMT with several lattice spacings ($\beta =2.1, 2.2, 2.3, 2.4, 2.5$).
The fits with two fitting forms are also displayed.
The colored band shows the uncertainty.
The data points at $r=0,1$ (lattice unit) were removed since they are contact terms.
Those at $r \ge 4$ were also erased since the error bars are too large, thus negligible in the fit.
}
\end{figure}

The result of our work is plotted in Fig.~\ref{fig:potential}.
In the fit of the potential, we removed the data points at $r=0$ and $r=1$ (in lattice unit) since they  are considered as contact terms (note that the glueball operator has a spread of one lattice unit).
The important feature of our result is that the interglueball potential is attractive.
We fit the potential with two fitting forms, i.e. with the Yukawa function $V_Y(r) = V_Y^{(1)} \frac{e^{-m_\phi r}}{4 \pi r}$ and with two Gaussians $V_{G}(r) = V_G^{(1)} e^{-\frac{(m_\phi r)^2}{8}} +V_G^{(2)} e^{-\frac{(m_\phi r)^2}{2}}$.
Afterthen we find $V_Y(r) = -231(8) \frac{e^{-m_\phi r}}{4 \pi r}$ ($\chi^2/$d.o.f. = 1.3) and $V_{G}(r) = -8.5(0.5) \Lambda e^{-\frac{(m_\phi r)^2}{8}} -26.6(2.6) \Lambda e^{-\frac{(m_\phi r)^2}{2}}$ ($\chi^2/$d.o.f. = 0.9).
Our extracted Yukawa coupling $V_Y^{(1)}$ has the same order of magnitude as that extracted in lattice $SU(3)$ YMT using L\"{u}scher's finite volume method \cite{Luscher:1990ux}, $V_Y^{(1)}=-650 \pm 190$ \cite{deForcrand:1984eeq}. 
It is also notable that it is comparable to the one-pion exchange nuclear force in QCD, namely $V_{\pi N} = -\frac{g_A^2 m_N^2}{f_\pi^2} \approx - 160$.
By matching the tree level one-glueball exchange process and our fitted $V_Y(r)$, the two low energy constants of the glueball effective Lagrangian (\ref{eq:glueballEFT}) are obtained as
\begin{eqnarray}
|f_\phi | 
&\approx &
1.4 \Lambda
,
\\
H_0 
&\approx &
-18 \Lambda^4
,
\end{eqnarray}
with a theoretical uncertainty of about 50\%.

Now that we have the analytic form of the potential, we calculate the scattering phase shift and the cross section.
The scattering phase shift is obtained by simply solving the following (s-wave) Schr\"{o}dinger equation:
\begin{eqnarray}
\biggl[
\frac{\partial^2}{\partial r^2} 
+k^2
-m_\phi V(r)
\biggr]
\phi(r)
=0
.
\end{eqnarray}
The wave function asymptotically behaves as $\phi(r) \propto \sin [kr+\delta(k)]$, where $\delta(k)$ is the scattering phase shift.
In the context of the DM, we are interested in the (s-wave) low energy limit of the cross section $\sigma_{\phi \phi} =\lim_{k\to 0}\frac{4 \pi}{k^2} \sin^2 [\delta(k)]$.
From the two fitting forms, we obtain $\sigma_{\phi \phi} = (2.5 - 4.7) \Lambda^{-2} $ (Yukawa), and $\sigma_{\phi \phi} = (14 - 51) \Lambda^{-2} $ (2-Gaussian), with the band denoting the statistical error.
By considering the difference between them as the systematic error, the interglueball scattering cross section in the $SU(2)$ YMT is 
\begin{eqnarray}
\sigma_{\phi \phi} 
= 
(2 - 51) \Lambda^{-2} \ \ \ {\rm (stat.+sys.)}.
\label{eq:gblimit}
\end{eqnarray}

Let us now derive the constraint on $\Lambda$ from observational data.
The most robust bound on the DM cross section is given by the observation of the galactic collisions, $\sigma_{\rm DM} / m_{\rm DM} < 0.47 \, {\rm cm }^2 /$g \cite{Harvey:2015hha}.
By substituting our result (\ref{eq:gblimit}), we obtain 
\begin{eqnarray}
\Lambda
>
60
\,{\rm MeV}
.
\end{eqnarray}
To our knowledge, this is the first quantitative constraint on the dark YMT derived from lattice gauge theory.

Let us add several comments on these results.
If we also consider the discussion of Spergel and Steinhardt which derived a lower limit on the DM cross section $\sigma_{\rm DM} / m_{\rm DM} > 0.45 \, {\rm cm }^2 /$g \cite{Spergel:1999mh}, we can almost determine the scale parameter of the YMT.
As seen in the beginning, this bound could be set by inspecting that the so-called core-vs-cusp, too-big-to-fail, and missing satellite problems could be resolved by assuming a finite DM scattering cross section.
There are however alternative possibilities to resolve the above-mentioned problems \cite{Chan:2015tna,Wetzel:2016wro,Kim:2017iwr,Bullock:2017xww}, so we leave the lower limit on $\Lambda$ open and wait for a conclusive resolution.

We also note that the glueball effective Lagrangian (\ref{eq:glueballEFT}) that we determined in this work has a wide range of applicability in cosmology.
It is actually known that in YMT the deconfinement transition occurs at $T_c \sim 1 \times \Lambda$ \cite{Lucini:2012gg}, which is much less than the glueball mass $m_\phi \sim 6 \Lambda$.
The glueball effective Lagrangian may thus be used even just below $T_c$, and this will permit us to calculate other important cosmological observables such as the DM relic density with reasonable accuracy.
This means that we now have in our hand a well justified and precise framework to calculate the cosmology in dark $SU(2)$ YMT.

Our discussion may be extended to larger $N_c$'s.
The YMT's with $N_c\ge 3$ are indeed even more interesting as candidates of dark sectors since the deconfinement transition at finite temperature is of the first order \cite{Svetitsky:1982gs,Lucini:2012gg}, and they may be probed with the background gravitational waves \cite{Kamionkowski:1993fg,Maggiore:1999vm,Schwaller:2015tja,Caprini:2015zlo}.
We can qualitatively estimate the limits on their scale parameters according to the large $N_c$ argument.
Since the cross section scales as $1/N_c^4$, we can derive the lower limits on the scale parameter as
\begin{eqnarray}
\Lambda_{N_c}
>
60 
\left( \frac{2}{N_c} \right)^{\frac{4}{3}}
\,{\rm MeV}
.
\end{eqnarray}
Of course, $1/N_c$ corrections are not small for $N_c =2$, so the first principle calculations on lattice have to be definitely done for $N_c \ge 3$ until they become sufficiently small.

\section*{Acknowledgements} 
This work was supported by ``Joint Usage/Research Center for Interdisciplinary Large-scale Information Infrastructures'' (JHPCN) in Japan (Project ID: jh180058-NAH).
It was also partially supported by Grant No. 0657-2020-0015 of the Ministry of Science and Higher Education of Russia.
The calculations were carried out on SX-ACE at RCNP/CMC of Osaka University.
MW was supported by the National Research Foundation of Korea (NRF) grant funded by the Korea government (MSIT) (No.~2018R1A5A1025563 and No. 2019R1A2C1005697).

\end{document}